\documentclass[10pt, openany]{book}

\usepackage[]{graphicx}
\usepackage{booktabs}
\usepackage[]{color}
\usepackage{alltt}
\usepackage[T1]{fontenc}
\usepackage{adjustbox}
\usepackage[utf8]{inputenc}
\usepackage{threeparttable}
\usepackage{graphicx}
\usepackage{amsmath}
\usepackage{lipsum}
\usepackage{color}
\usepackage{rotating}
\usepackage{multirow}
\usepackage{hhline}
\usepackage{wrapfig}
\usepackage{caption}
\usepackage{hyperref}
\usepackage{cite}
\usepackage[table,xcdraw]{xcolor}

\usepackage{tabularx}

\usepackage[ruled,vlined]{algorithm2e}
\usepackage{algorithmic}

\usepackage{relsize}
\def\textsubscript#1{\ensuremath{_{\mbox{\textscale{.6}{#1}}}}}

\usepackage[top=100pt,bottom=100pt,left=68pt,right=66pt]{geometry}

\usepackage{lipsum}

\usepackage[swedish, english]{babel}

\usepackage{fancyhdr}
\usepackage{titlesec}
\titleformat{\chapter}
   {\normalfont\LARGE\bfseries}{\thechapter.}{1em}{}
\titlespacing{\chapter}{0pt}{50pt}{2\baselineskip}

\usepackage{float}
\floatstyle{plaintop}
\restylefloat{table}

\usepackage[tableposition=top]{caption}

\usepackage{hyperref}
\hypersetup{hidelinks,linkcolor = black} 

\graphicspath{ {images/} }

\frontmatter

\begin{document}


\begin{titlepage}
	\clearpage\thispagestyle{empty}
	\centering
	\vspace{2cm}

	{\large Wireless/Mobile Multimedia Networks - COEN 332 \par}
	\vspace{4cm}
	{\Huge \textbf{Point to Point Ethernet Transmission Wireless
Backhaul Links Clustering}} \\
	\vspace{1cm}
	{\large \textbf{} \par}
	\vspace{4cm}
	{\normalsize Puneet Kumar \\ 
	             SCU-ID: 00001424550 \par}
	\vspace{2cm}

    \includegraphics[scale=0.60]{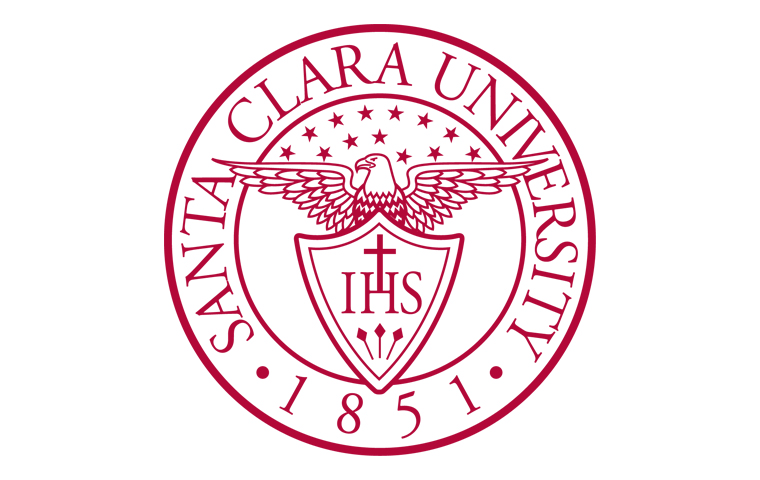}
    
    \vspace{2cm}
    
	{\normalsize Santa Clara University \\ 
		Computer Networks: COEN-332 \\
		Department of Computer Science \par}
		
	{\normalsize California, USA \par}
	\vspace{2cm}
	
	\pagebreak

\end{titlepage}

\chapter{Audience}
\label{chapter_1}
\large This document is intended to explain the Wireless Backhaul Links Clustering between point to point ethernet transmission links. 
With the word clustering the first thing comes to
mind is bundling bunch of physical links to make a single
logical link. While this definition holds true, it is imperative
to know it’s implementation in conjunction with certain key
factors. Upcoming 5G technologies where wireless backhaul
links for point to point wireless Ethernet transmission may
play a vital role in backbone networks, the proliferated needs
of providing higher data rates should be met. With that
demand, challenge of resilience, performance, scalability,
maintainability, and manageability will rise as well. An
optimal solution must be found which can reduce the failover
points, ability to perform upgrade and downgrade with
minimal downtime, and most importantly solution must
overcome all the challenges faced in wireless link
environment. A well tested approach has been presented in
this paper which will accomplish all the mentioned points.
This method will make the clustered wireless backhaul links
more valuable and throughput will increase significantly.

This document targets audience of the class who are taking taken COEN - 332 (Wireless/Mobile Multimedia Networks), as fundamentals to understand this document was explained in the class. 
Additionally, lectures coverd layer 2/3 and wireless protocols for control, management and data plane required for this project report.

\tableofcontents{}
\listoffigures
\listoftables

\mainmatter

\chapter{Introduction}
\label{intro}
Wireless Ethernet Transmission is a great way of
extending the backbone network without deploying any
physical network, specially fiber. This technology can be
deployed in any form of topology and network. After FCC
rolled out extra free frequency bands for instance 11GHz
and 5GHz for commercial use, its demand is on rise \cite{goovaerts2016fcc}.
Today Internet Service Providers (ISP), Enterprise
Networking companies are relying on this technology a lot
due to numerous factors such as Cost, Terrains,
maintenance, easy to upgrade, and most importantly easy
to troubleshoot.
With upcoming 5G technologies where connecting 5G
access points to the backbone networks will require to have
a relay link in between to keep the data rate constant for
long links. Wireless Ethernet Transmission technology
will be handy to serve as a middleman in between
backbone networks and 5G access points, where the link is
just too long and data rate most likely will suffer without a
Wireless Ethernet Transmission backhaul link. Another
application of Wireless Transmission Network is campus.
In campus, often it is hard to deploy fiber to extend the
network to other buildings. Since it is in short range, so the
interference wouldn’t matter that much. These backhaul
links comes handy. Fiber gets dropped by the ISP to some places and then the network gets extended by these
backhaul links.

\begin{figure}[t]
    \centering
    \includegraphics[width=0.6\linewidth]{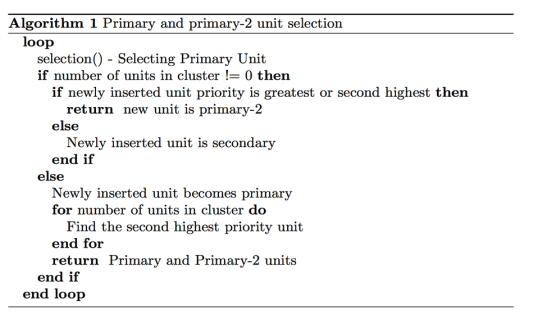}
    \captionsetup{font=footnotesize}
    \caption{Primary and Primary-2 selections
    }
    \label{fig:primaryPrimary2}
\end{figure}

In all these above-mentioned applications, we would
need a higher amount of data rate with reliability and
ability to be up for maximum amount of time. With the
help of current MIMO technologies, specially with IEEE
802.11ac, if we have 2 RF chains, 4 Streams then the max
data rate (Consider optimal RF conditions) is around
1Gbps \cite{Mimosa_B5}. With such high demand of data rate, 1Gbps
will not be sufficient and hence will not be a suitable
solution after all. In 5G the data rates will be way more
than 1Gbps, deploying Wireless Ethernet Transmission
Backhaul links in 5G will not suffice the purpose. An
optimal solution is needed to make data rates higher with
the existing infrastructure. Cluster will bundle multiple
Wireless Ethernet Transmission Backhaul Links and will
become a logically single link to provide higher data rates
needed \cite{802_1X}. There will be a lot more to be explained and
discussed with clustering, few are outlined below:
\begin{itemize}
    \item Control Plane messages in unit selections and
assigning duties
    \item Load Balancing across links
    \item High Availability
    \item Flow consistency
    \item State replication
    \item Configuration replication
    \item Management plane, Control plane, Data plane traffic
handling
\end{itemize}

This solution will provide multiple Gbps data rate but also
will provide a scalable, resilient, highly available, easy to
upgrade with minimal disruption solution.

\chapter{Concepts and Explanation}
\large  This section will explain the concepts and their relevance in clustering in wireless backhaul links.
\begin{figure}[t]
    \centering
    \includegraphics[width=0.5\linewidth]{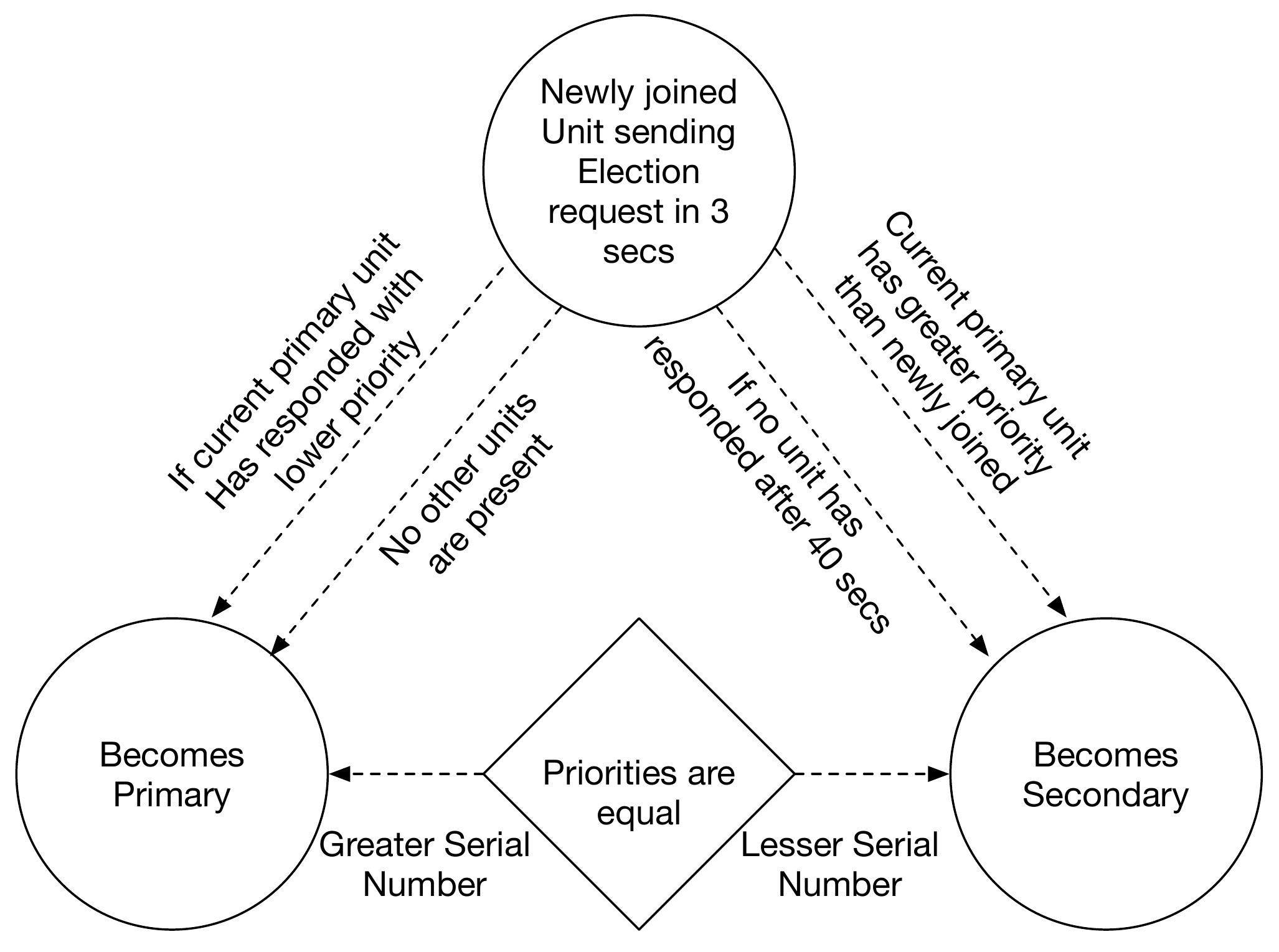}
    \captionsetup{font=footnotesize}
    \caption{Newly joined unit selection process 
    }
    \label{fig:newlyJoinedUnit}
\end{figure}

\begin{figure}[t]
    \centering
    \includegraphics[width=0.7\linewidth]{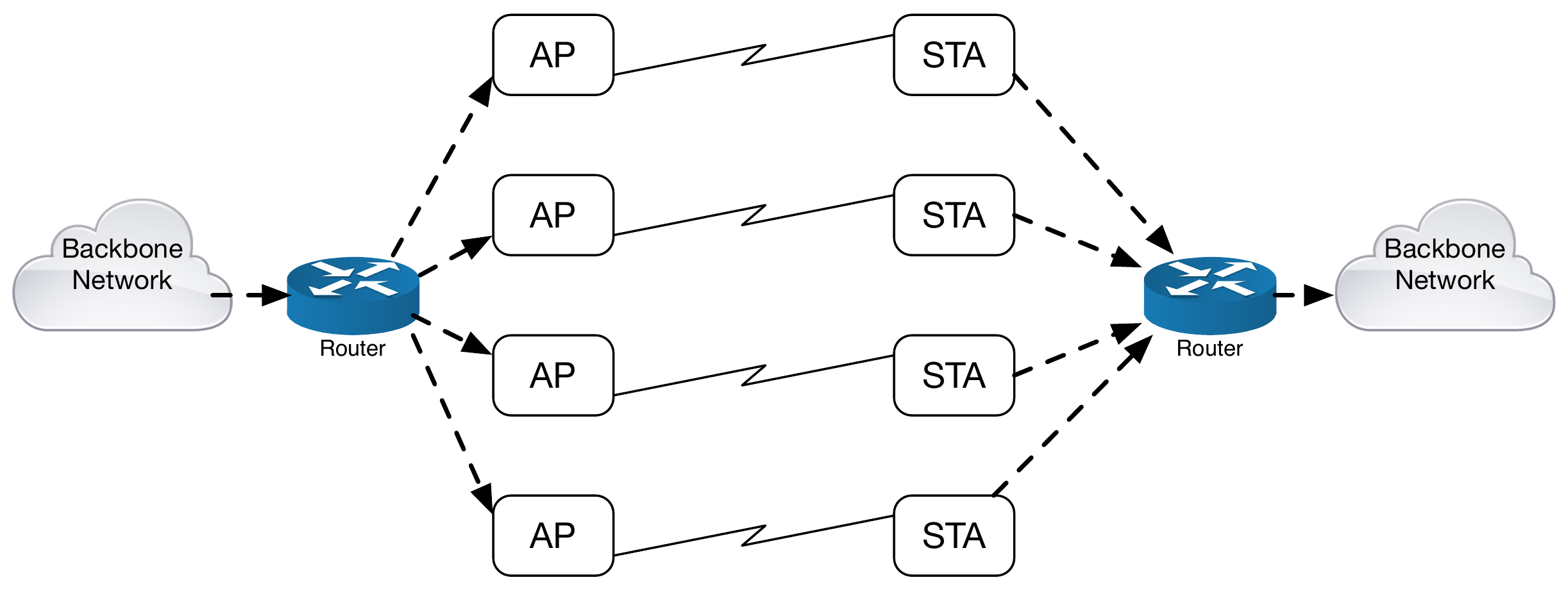}
    \captionsetup{font=footnotesize}
    \caption{Physical Topology. AP = Access Point, STA = Station
    }
    \label{fig:physical_topology}
\end{figure}

\begin{figure}[t]
    \centering
    \includegraphics[width=0.8\linewidth]{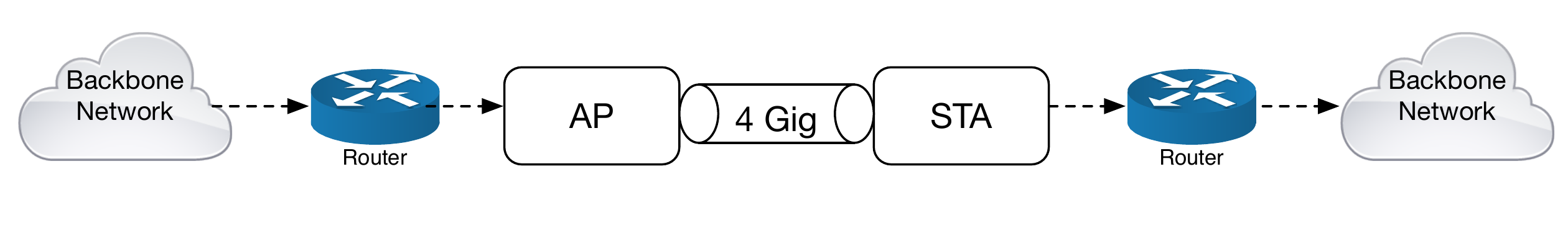}
    \captionsetup{font=footnotesize}
    \caption{Logical Topology. AP = Access Point, STA = Station
    }
    \label{fig:logical_topology}
\end{figure}

\section{Cluster Formation} 
Cluster will be a daemon running on each radio. Which
will bind all the radios wireless links. This will appear one
link for south and north bound devices and will aggregate
the throughput. Cluster will have failover mechanism, load
balancing and an algorithm for communication. Topology
is needed to understand it as shown in figure \ref{fig:physical_topology} and figure \ref{fig:logical_topology} which
shows the difference between physical topology and the
logical topology. 

\section{Bootstrap Configuration and Cluster Members} 
A unit consists of access-point and station. Each unit
requires minimal bootstrap configuration which includes
cluster name, cluster control link interface, management
cluster link interfaces, data path link interfaces and local
pool of IP address reserved for clustering.
The first unit configured as a bootstrap configuration will
be primary unit and rest become “secondaries”. These are
initial roles; The primary unit assignment will also depend
on the priority set in the bootstrap configuration. Priority
can go from 1-100, where 1 is the highest priority. All
other members are secondary units except one unit with
second highest priority will be primary-2 and secondary
both. Algorithm (figure \ref{fig:primaryPrimary2}) shows the complete scenario. In typical scenario, the very first unit added becomes
the primary, it is because so far this is the only unit present
in the cluster. Apart from bootstrap configuration, all
configuration is on primary only; replication of
configuration on secondary unit starts. When it comes to
physical assets, for instance interfaces, primary’s
configuration is mirrored to secondary unit. For example,
if Ethernet 1 is configured as inside and Ethernet 2 is
configured as outside, then these interfaces on secondary
unit will be inside and outside respectively.
Primary unit selection is based on request message which
consists of priority set by user. As soon as a new unit is
joined in the cluster, an election request will be sent on a
multicast address 224.1.0.10 which is reserved for this
purpose. All units in cluster will listen to this IP and only
primary unit at that moment will reply to the newly joined
unit. If the priority of newly joined unit is greater than the
primary unit then newly joined unit will set itself to
primary (Please see figure \ref{fig:newlyJoinedUnit}). If newly joined unit doesn’t receive any response
back from primary unit then it sets itself and sends a
multicast message to 224.1.0.11 which will force every
unit to set them to secondary. In the case of tie of priorities
(especially the highest priority), different parameters such
as unit name and serial number are used to determine the primary unless it is down or stops responding for some reason, this will trigger a new unit selection 

\begin{figure}[t]
    \centering
    \includegraphics[width=0.8\linewidth]{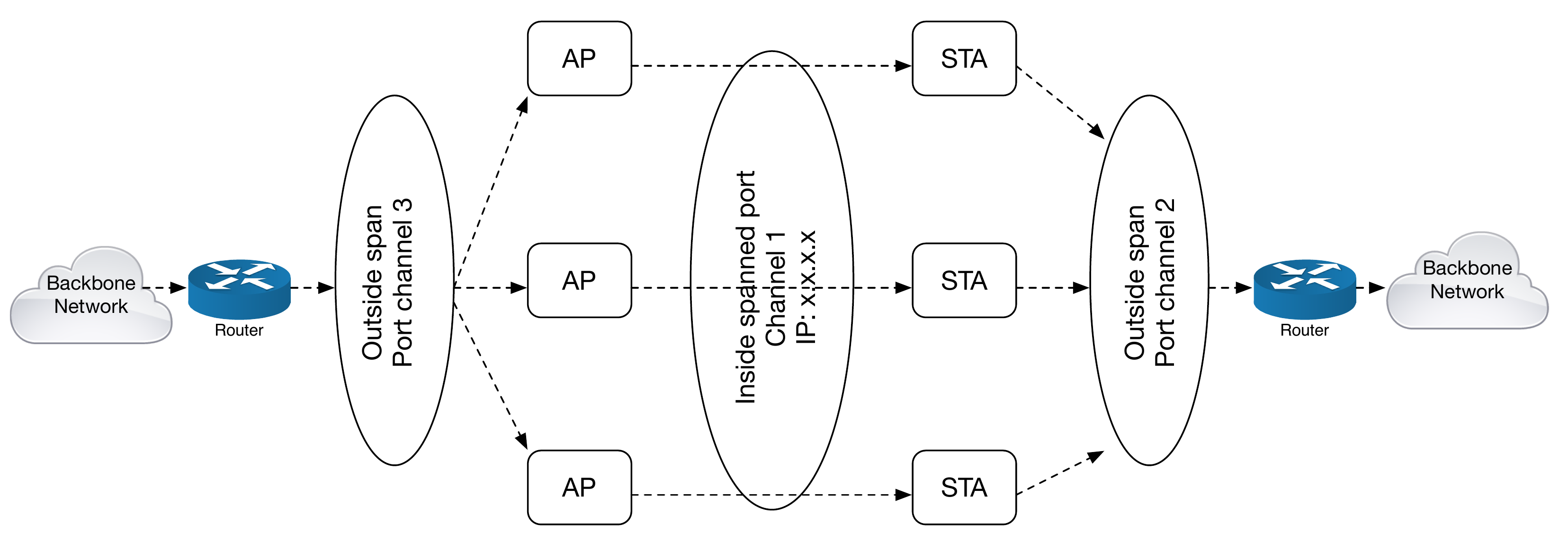}
    \captionsetup{font=footnotesize}
    \caption{Spanned EtherChannel with LACP
    }
    \label{fig:SpannedEtherChannel}
\end{figure}

\begin{figure}[t]
    \centering
    \includegraphics[width=0.8\linewidth]{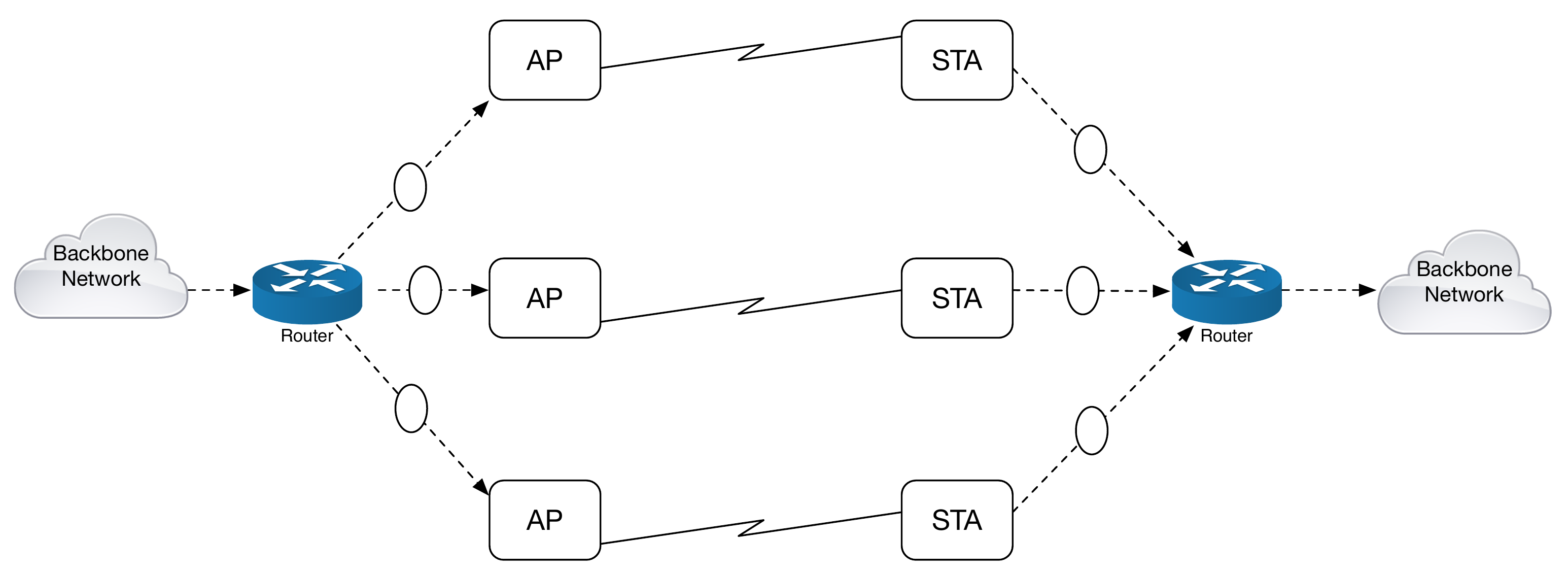}
    \captionsetup{font=footnotesize}
    \caption{ Individual router interface mode
    }
    \label{fig:IndividualRouterInterfaceMode}
\end{figure}

\section{Cluster Interfaces}
All the wireless backhaul links can either be configured
as Spanned EtherChannel or Individual interfaces. All
links has the configured one type only. Spanned
EtherChannel is recommended. One or more links can be
grouped into Spanned EtherChannel that spans all links in
the cluster. Both routed and transparent modes are present
in Spanned EtherChannel. A single IP address is provided
to routed interface if the EtherChannel is configured in
routed mode, on the other hand, transparent mode is
something when we have an IP assigned to BVI (Bridge
virtual Interface) not to the member interface of bridge
group. This will inherit the load balancing as a part of basic
operation.
In Spanned EtherChannel shows in figure \ref{fig:SpannedEtherChannel}, we will
combine all the interfaces into a single logical link using
LACP. This interface will become a bundle of all the
interfaces attached to it. This EtherChannel will spans all
units in the cluster. It is recommended as EtherChannel
aggregates traffic from all the available active interfaces in
the channel. In this mode, all units use the same VIP
(Virtual Internet Protocol address) assigned to the cluster
and the same MAC used in EtherChannel of the cluster.
Another mode is individual routed mode (Figure \ref{fig:IndividualRouterInterfaceMode}) where each link
will maintain its own IP on each data interface. Since this
approach doesn’t use any aggregation link to load balance
and implement any fail over, dynamic routing is needed to
load balance. In individual mode, each radio maintains its
own routing adjacency. The disadvantage of this is slower
convergence and higher processor utilization due to each
unit maintaining its own routing table. In spanned
EtherChannel mode, the primary access point and station
runs dynamic routing. Routing and ARP tables are
synchronized to the secondary units.
A round trip time of 20ms and maintaining a reliable
cluster control link functionality is important. To make
sure that this condition is intact, a ping pong icmp
messages goes from every secondary units to master units.
Cluster control link is the most important link in the cluster
and will have a dedicated interface for cluster control link.
If a cluster control link is down, then the cluster will be
down and it must be brought up by solving the problem
which led it to down and then joining them one by one
manually.

\section{Monitoring units }
\begin{figure}[t]
    \centering
    \includegraphics[width=0.8\linewidth]{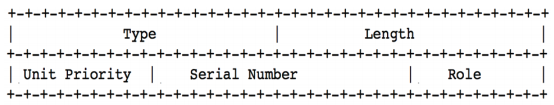}
    \captionsetup{font=footnotesize}
    \caption{Selection Info Component  
    }
    \label{fig:SelectionInfoComponent}
\end{figure}

\begin{figure}[t]
    \centering
    \includegraphics[width=0.6\linewidth]{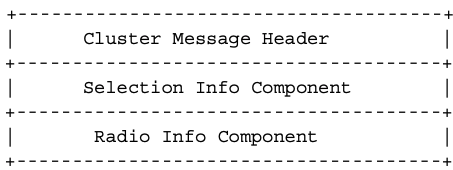}
    \captionsetup{font=footnotesize}
    \caption{Keepalive Message format 
    }
    \label{fig:KeepAliveMessageFormat }
\end{figure}

To make sure all units are healthy and interfaces are up
and running, a monitoring system is placed. Primary unit
is responsible to monitor every secondary unit by sending
a keepalive message periodically over the CCL (Cluster
Control Link), this monitoring period is configurable. In the event of unit health check failure, it will be removed
from the cluster. Every unit is supposed to send a keepalive message
(UDP multicast packet) (Figure \ref{fig:KeepAliveMessageFormat } to primary unit. Message format: 

\begin{itemize}
    \item   Keepalive message header
        \begin{itemize}
             \item   Type 
             \item Version
             \item Length
        \end{itemize}
        
    \item Selection Info Component
            \begin{itemize}
                 \item Unit Priority
                 \item Serial Number
                 \item Role (Primary Standby or Secondaries)
            \end{itemize}
            
    \item   Radio Info Component
            \begin{itemize}
                 \item Mode (Spanned EtherChannel or Individual Router Interface
                 \item Radio Type (Access Point or Station)
                 \item SNR
                 \item Load Balancing Weight
            \end{itemize}
\end{itemize}
Here type is the keepalive (CLUSTER\_KEEPALIVE).
Version is 1 and length is the length of the message not
including the header. The message has two more sections
apart from the header. Section info component and Radio
Info component. 
Figure \ref{fig:SelectionInfoComponent} shows the message header. Here type is the SELECTION\_INFO\_COMP, length is
the length of the selection info component. Unit priority is
the priority assigned to unit while configuring cluster.
Serial Number of the unit and Role of the unit whether it
is Primary standby or a secondary unit. This section will
let primary unit know about configurations of other units.

\begin{figure}[t]
    \centering
    \includegraphics[width=0.8\linewidth]{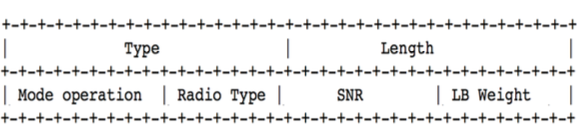}
    \captionsetup{font=footnotesize}
    \caption{Radio Info Component 
    }
    \label{fig:RadioInfoComponent}
\end{figure}

\begin{figure}[t]
    \centering
    \includegraphics[width=0.8\linewidth]{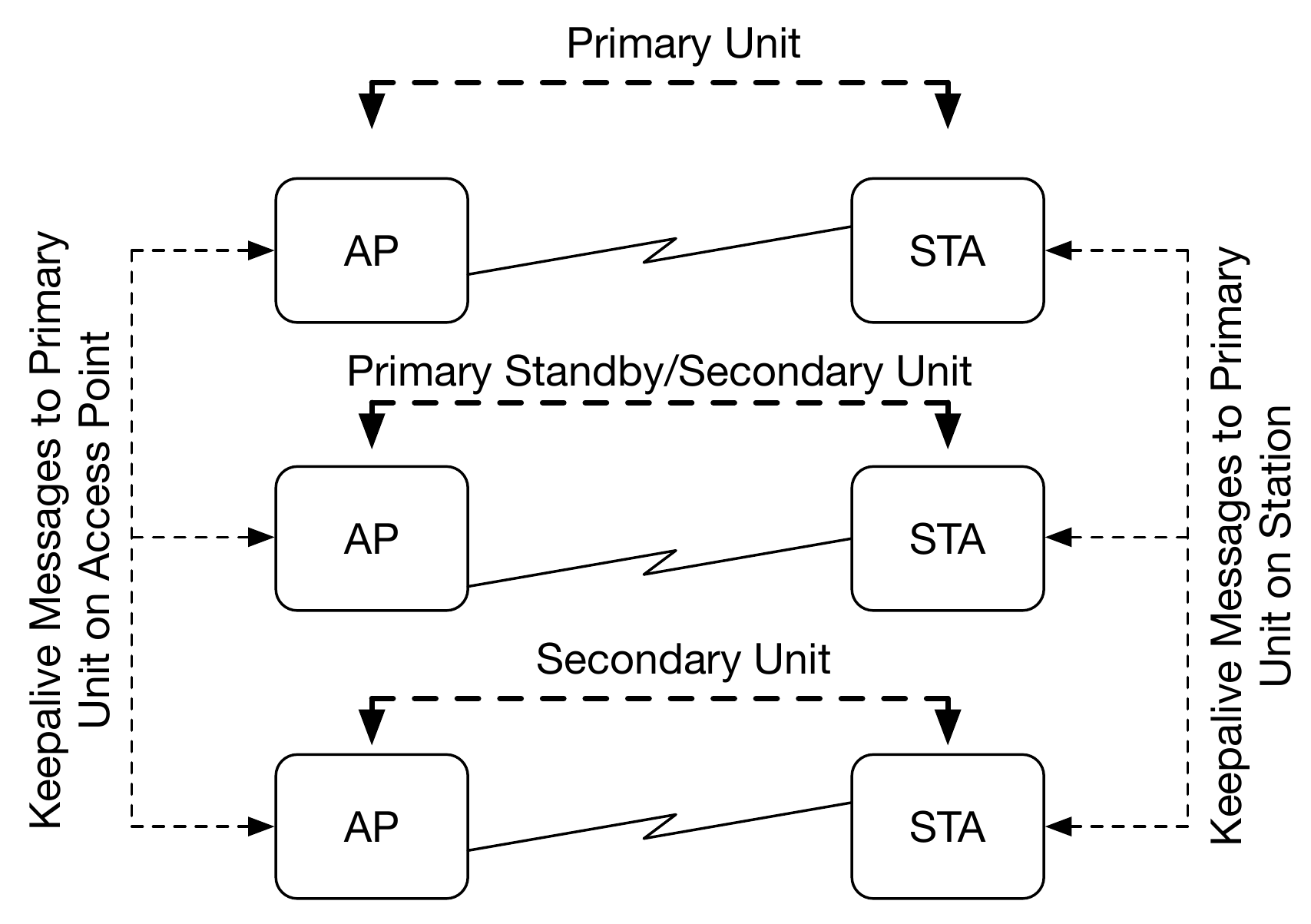}
    \captionsetup{font=footnotesize}
    \caption{Message Flow to Primary Unit
    }
    \label{fig:MessageFlowtoPrimaryUnit }
\end{figure}

Figure \ref{fig:RadioInfoComponent} shows the radio info component message header.
Here the type is RADIO\_INFO\_COMP.
Length is the length of radio info component. It includes
Mode operation whether it is spanned EtherChannel or
Individual router interface. If primary unit finds that
received packet has mismatch of mode of operation, then
it excludes the unit from cluster and sends a unicast
message to forcefully make the unit leave cluster. A high level message flow is depicted in figure \ref{fig:MessageFlowtoPrimaryUnit }.

\section{Monitoring Interfaces and failure status}
As soon as the health monitoring is turned on, all physical
interfaces are monitored, including the main EtherChannel
and redundant interface. An option will be provided to user
to disable the health monitoring per interface. For instance, an EtherChannel is considered to be failed if all the units
in the EtherChannel are failed. If this happens, the
EtherChannel will be removed from the cluster but this
will depend on the minimal port bundling settings. For a
single unit, it will be considered removing from the cluster
if all monitored interfaces fail. The amount of time
removing a unit from the cluster will depend on the type of
interface and whether a unit is established or just joining
the cluster. For any EtherChannel spanned or not, if an
interface is down on an established member, primary unit
removes it after 9 seconds. For first 90 seconds, primary
unit doesn’t monitor any interfaces for the unit which just
joined the cluster. This means a primary unit will not
remove it if the interface state changes during that 90 secs.
In non-EtherChannel, the unit is removed in 500 ms
regardless of the member state. In the case of failure of a
unit, the connection belongs to that unit is seamlessly
transferred to other units and state information for the
traffic flows are shared over the CCL (Cluster Control
Link). In case of primary unit failure itself, other member
with highest priority (Which means unit with the lowest
number priority) takes over. Failed primary unit, after
recovery automatically tries to join the cluster. If rejoining
the cluster of primary unit is failed then all data interfaces
are shut down and only management interface can receive
and send traffic. This management interface remains up
using the IP address the unit received from the cluster IP
pool.

\section{Cluster Rejoining}
The primary factor to join back a unit after being
removed from the cluster is the reason why the unit was
removed. There are three methods how a unit can rejoin
the cluster. 
\begin{itemize}
    \item Failed cluster control link when initially
joining—After the problem is resolved with the
cluster control link, unit must be manually
rejoined the cluster by re-enabling clustering at
the console port by entering cluster group name,
and then enable. 
     \item Failed cluster control link after joining the
cluster—Unit automatically tries to rejoin every
5 minutes, indefinitely. This behavior is
configurable. 
     \item Data Interface Failure - Unit will try maximum of
4 attempts on each 5 minute to rejoin the cluster.
After resolving the issue with the data interface,
a manually enabled clustering is required by
entering cluster group name. This behavior is
configurable. 
\end{itemize}

\begin{figure}[t]
    \centering
    \includegraphics[width=0.8\linewidth]{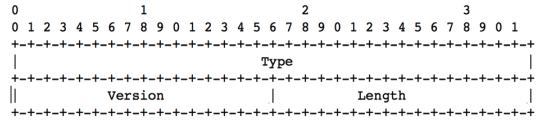}
    \captionsetup{font=footnotesize}
    \caption{Cluster message header.  
    }
    \label{fig:ClusterMessageHeader}
\end{figure}

\begin{table}[h]
\caption{Extra information in connection preservation}
\begin{center}
\begin{tabular}{|c|c|c|}
\hline
\label{SampleTable}
Traffic           & State Support & Nodes                                                                                                         \\ \hline
Up Time           & Yes           & \begin{tabular}[c]{@{}c@{}}Keeps track of the \\ system up time\end{tabular}                                  \\ \hline
ARP Table         & Yes           & \begin{tabular}[c]{@{}c@{}}Individual interface \\ routed mode only\end{tabular}                              \\ \hline
MAC Address Table & Yes           & \begin{tabular}[c]{@{}c@{}}Individual Interface \\ Router Mode Only\end{tabular}                              \\ \hline
User Identity     & Yes           & \begin{tabular}[c]{@{}c@{}}Includes AAA (authentication)\\ and radius server related information\end{tabular} \\ \hline
SNMP Engine ID    & No            & ------------------------------                                                                                \\ \hline
VPN Site-Site     & No            & ------------------------------                                                                  \\ \hline
\end{tabular}
\end{center}
\end{table}

\section{Replication of Data path connection state}
\begin{figure}[t]
    \centering
    \includegraphics[width=0.5\linewidth]{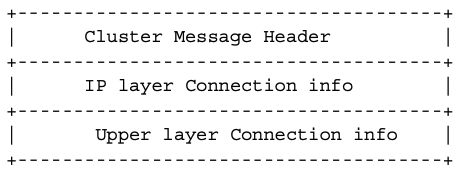}
    \captionsetup{font=footnotesize}
    \caption{Replication message format 
    }
    \label{fig:ReplicationMessageFormat }
\end{figure}

\begin{figure}[t]
    \centering
    \includegraphics[width=0.8\linewidth]{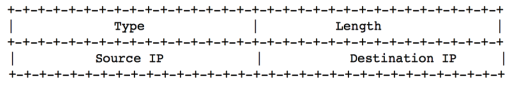}
    \captionsetup{font=footnotesize}
    \caption{IP Layer connection info 
    }
    \label{fig:IPLayerConnectionInfo }
\end{figure}

There is one owner and one backup owner for every
connection in the cluster. Ownership doesn’t get
transferred to backup owner in case of a failure instead a
provision of TCP/UDP state information gets stored in it
so that the connections can be transferred seamlessly. If for
some reason the owner becomes unavailable, the very first
unit to receive packets from the connection contacts the
backup owner for the relevant connection state and then the backup owner becomes a new owner for this
connection. Obviously, there would be some traffic which
would require information above the TCP or UDP layer.
Replication message is defined below. Cluster message
header is already defined in figure \ref{fig:ClusterMessageHeader}. In the type field of
figure \ref{fig:ReplicationMessageFormat }, type field will be mentioned in
CLUSTER\_REPLICATION, version will be 1 and length
is the length of the packet apart from the cluster message
header. 
Often the traffic will be directed by the original unit
where the traffic is flowing. Often UDP/TCP or higher
layer information not needed to be transferred to the
backup unit or in case of transfer to back up unit the traffic
could be from the same origin and destined to the same
destination. Separating the IP layer info (message shown in figure \ref{fig:IPLayerConnectionInfo } from the higher
layer can give flexibility to just check the source and
destination IP and not needed to open the rest of the packet.
This will make the transfer faster.
Upper layer connection info (Message shown in figure \ref{fig:UpperLayerConnectionInfo} will contain vital
information about Upper layer as shown below. Most of
the time, the information doesn’t change and failover will
only have to see the layer-3 packet and forward the upper
layer information. Also, table \ref{SampleTable} shows what extra
information in preserved in connection and its transfer. 

\section{Cluster Management}
To manage the cluster, a separate network must be
created apart from cluster control network. A spanned
EtherChannel or individual interfaces can be used as
management network. Even if spanned EtherChannel is
used for data still for management purposes recommended
approach is individual interface. In it each unit can be
connected directly to individual interfaces (if necessary),
on the other hand a spanned EtherChannel interface only allows to be connected to primary unit via remote
connection.
In case of individual interface, cluster IP address is fixed
which always belongs to the current primary unit. A range
of IP address can be configured for each unit which of
course includes the current primary which can use a local
address from the range. Having a main cluster IP provides
management consistency access to an address; In case of
role change of primary unit to other unit in cluster, this
main IP moves to the new unit which will result in the
seamless transfer of the management of the cluster. For
routing, the local IP is used and it is also useful for
troubleshooting. For an instance, a cluster is managed by
the main cluster IP and this IP is always attached to the
current primary unit but to manage an individual member,
a local IP can do the job. There are outbound management
traffic service like syslog, TFTP etc, each unit (includes
primary unit too) uses this local IP to connect to the server.
In the event of Spanned EtherChannel interface, one main
IP is configured which is attached to the primary unit.
Secondary units will not be allowed to connect directly
using the EtherChannel interface and hence an individual
interface is recommended for configuring the management
interface and user will be easily able to connect to single
unit. 

\section{Load Balancing Method}
\begin{figure}[t]
    \centering
    \includegraphics[width=0.8\linewidth]{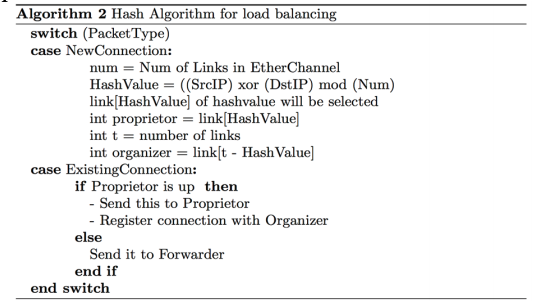}
    \captionsetup{font=footnotesize}
    \caption{Hash Algorithm for Load Balancing 
    }
    \label{fig:HashAlgo}
\end{figure}

Spanned EtherChannel have bundled links which load
balances and failovers in case of a link failure. There is
more than one physical links are bundled into the Spanned
EtherChannel (At Max 6), the primary aim is to have
traffic across all the links equally however a IP stickiness
is desired as well. Packets with same destination and
source IP should take the same path every single time for
IP stickiness to be maintained. This will give consistency
and ease of troubleshoot in case of data packet inspection
or loss. In the load balancing algorithm for EtherChannel,
a hash function (Algorithm shown in figure \ref{fig:HashAlgo} calculates a hash value which determines
which link packet will go out. The algorithm for hashing will be “symmetric” which
means the packet from both the directions will have the
same hash, and will be sent to the same unit in spanned 
EtherChannel. By default, Source IP and destination IP are
being used and it is recommended too. Another restriction
is to use same type of radio when connecting the units to
the switch so that hashing algorithm applied to all the
packets. In individual routed mode load balancing, each radio
will maintain its own IP address. One method of load
balancing is Policy Based Routing. Traditional Policy
based Routing is based on policy which is applied to
ingress and egress interface based on access-list which will
allow certain type of traffic to be passed. Policy is a map
which allows certain type of traffic to be passed from
certain units. Since it is static, chances are it may not
achieve the optimal load balancing results. Recommended
way of configuring policy is to make sure that forward and
return packets of a connection are directed to the same
physical unit. 

\section{Connection Management and Formation}
Roles of connections determine how they are handled in
high availability and normal operation. Connections can
be load balanced among multiple members of the clusters
too. For connection management, we will distribute the
unit roles of three different type:
\begin{itemize}
     \item Proprietor
     \item Organizer
     \item Forwarder
\end{itemize}
from the connection, the director chooses a new owner
from those units from the connection, the director chooses
a new owner from those units.
Function of Organizer is to handles owner lookup requests
which are coming from forwarders and maintain a connection state to serve as a backup if the owner fails.
When a Proprietor receives a new connection, based on our
hashing algorithm of source/destination IP a director is
chosen and a message is sent to the organizer to register
the new connection. If packet arrives at any unit other than
the owner then unit queries the director about which unit
is the owner so that forwarder can forward the packet to
the owner. A connection has only one Organizer. If an
Organizer fails, the owner chooses a new Organizer. Function of “Forwarder” is to forward packets to
Proprietor. If packet received by Forwarder doesn’t own
by it, it goes ahead and queries the organizer for the
proprietor and then it establishes a connection with the
Proprietor so that Forwarder can forward packets received
by it in future. An Organizer can also be a forwarder. Let’s
take an example, if a Forwarder receives a SYN-ACK
packet, it can derive the Proprietor directly from a SYN
cookie in the packet, so it does not need to query the
Organizer. There is a used case when TCP sequence
randomization is not enabled or disabled intentionally,
SYN Cookie is no longer useful and query to Organizer is
required. For short lived flows such as ICMP, DNS etc,
instead of querying, the forwarder immediately sends the
packet to the Organizer, which then sends them to the
Proprietor. A connection can have more than one
forwarder. A good load balancing algorithm is needed
where there are no forwarders and all packets of a
connection are received by the proprietor, our ECMP in
Spanned EtherChannel interface provides this. When a
new connection is directed to a member of the cluster via
load balancing, that unit owns both directions of the
connection. For any connection, if packets are arrived at a
different unit, they will be forwarded to the owner over
CCL (Cluster Control Link). If more optimization is
needed then an external load balancing needs to be in place
for both directions of the flow to arrive at the same unit, it
is redirected back to the original unit. 

\begin{figure}[t]
    \centering
    \includegraphics[width=0.7\linewidth]{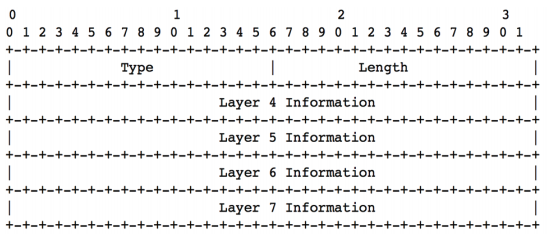}
    \captionsetup{font=footnotesize}
    \caption{Upper Layer Connection Info 
    }
    \label{fig:UpperLayerConnectionInfo}
\end{figure}

\begin{figure}[t]
    \centering
    \includegraphics[width=0.8\linewidth]{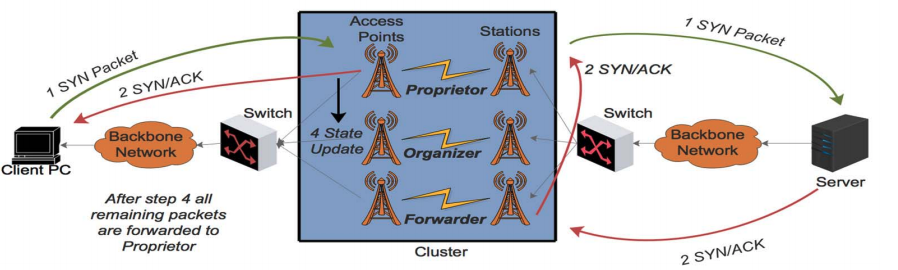}
    \captionsetup{font=footnotesize}
    \caption{Data Packet Flow 
    }
    \label{fig:DataPacketFlow}
\end{figure}

\section{Data Flow}
Figure \ref{fig:DataPacketFlow} shows the data packet flow.
Proprietor role is assigned to unit that initially receives
the connection. The owner maintains the TCP state and
processes packets. A connection has only one owner. If the
original owner fails, then when new units receive packets will become Proprietor.
\begin{itemize}
    \item The SYN packet originates from the client and it is
delivered to one Unit (based on the load balancing mechanism). This unit becomes the Proprietor,
proprietor creates the flow, encodes owner
information into a SYN cookie, and forwards the
packet to the server. 
    \item Since the forwarder doesn’t own the connection, it
uses SYN cookie to decode the owner information,
it creates a forwarding flow to the owner, and
forwards the SYN-ACK to the Proprietor. The
Proprietor sends a state update to the Organizer, and
forwards the SYN-ACK to the client. 
     \item  The Organizer receives the state update from the
Proprietor, creates a flow to the Proprietor, and
records the TCP state information as well as the
Proprietor. The Organizer acts as the backup owner
for the connection. 
 
     \item After this any subsequent packets delivered to the
forwarder will be forwarded to the proprietor. If
packets are delivered to any additional unit then a
query to director is needed for the owner and to
establish a flow. Any state change for the flow result
will be a state update from the proprietor to
organizer. 

    \item Unbalanced flow distributions resulted from load
balancing capabilities of the upstream and
downstream network can be mitigated by redirect
new TCP flows to the other units, this can be
configured while no existing flows will be moved to
the other units.
\end{itemize}

\chapter{Simulation Model}
\begin{figure}[t]
    \centering
    \includegraphics[width=0.65\linewidth]{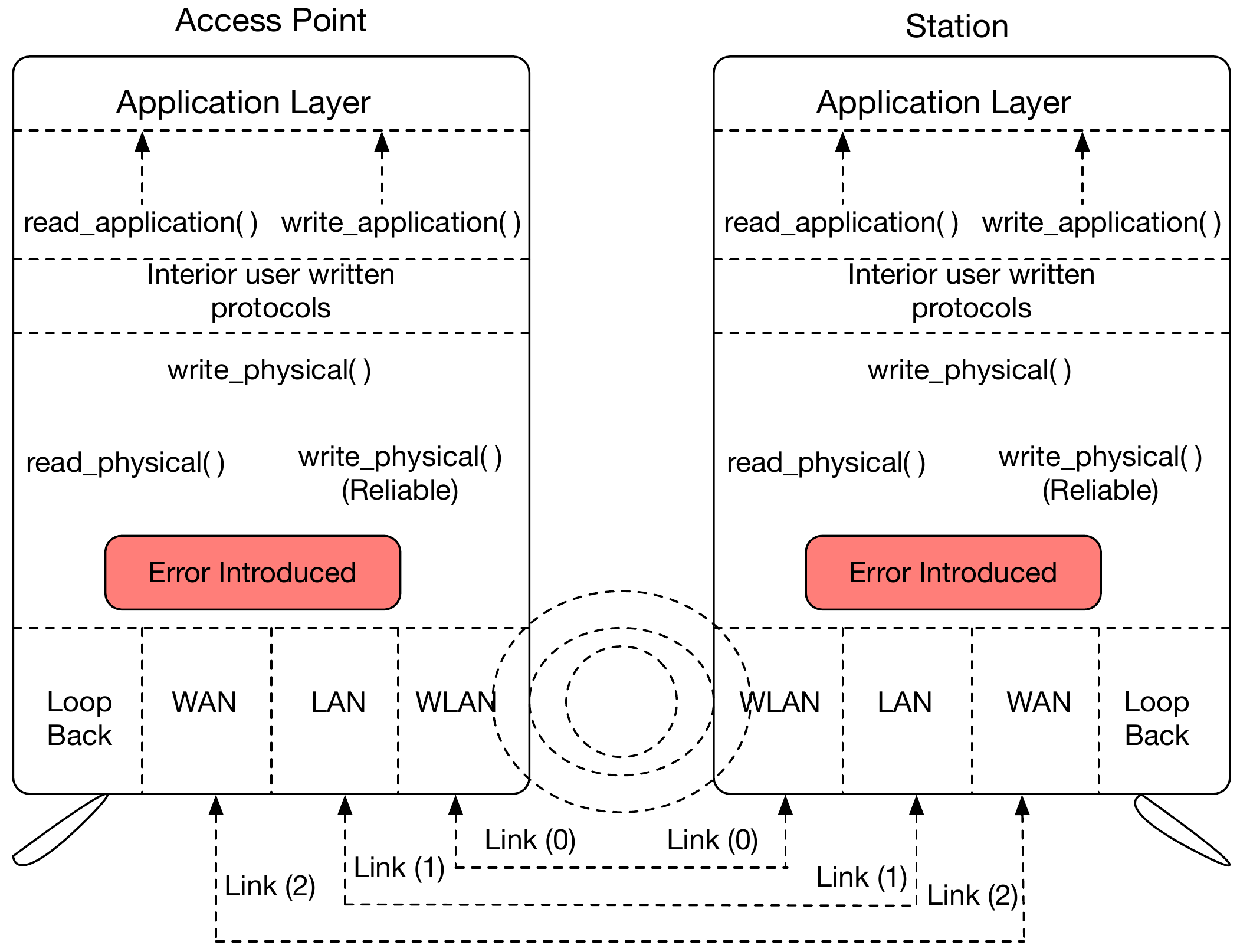}
    \captionsetup{font=footnotesize}
    \caption{Simulation model
    }
    \label{fig:simulation-model}
\end{figure}

We simulated wireless backhaul link by using CNET\cite{Simulator} (shown in figure \ref{fig:simulation-model}.
CNET use 5Ghz bandwidth.
CNET provides APIs (predefined) to set up its own simulation model.
We used customized parameters below for simulation.
\begin{itemize}
    \item 1472Bytes - MTU Size 
    \item 1 - 2 Gbps Bandwidth
    \item 5GHz - WLAN Frequency
    \item 20dBm - WLAN Tx Power
    \item 10dBi - WLAN Tx Antenna Gain
    \item 10dBi - WLAN Rx Antenna Gain
\end{itemize}

These tests are simulated while keeping some
parameters constant.
\begin{itemize}
    \item 80Mhz - Channel Width
    \item 2 (Number of Streams - 4) - No. of channels
    \item 30dBm - Tx Power
    \item 36dB - Signal to Noise Ratio (SNR)
    \item 0.5\% - Packet Error Rate (PER)
    \item (-17) - Error Vector Magnitude (EVM)
    \item 8 - MCS Index
    \item 256QAM - Modulation Technique
\end{itemize}

PHY rates demonstrated how much ethernet frames can propagate through the wireless backhaul links in optimal conditions.
Additionally, MAC rates are more fair comparison to the TCP throughput than PHY rates.
We already mentioned in previous section about how MAC rates are calculated.
After changing PHY rates, we will note how MAC rates are changing as well.
This will demonstrate the real world scenario about MAC rates, depends on different RF condition.

We used iperf for performing TCP throughput test.
Iperf gives us mechanism to customzie our TCP parameters.
We will set our iperf
parameters as: number of TCP Connection – 50 and TCP Window size - 64K.
PHY Rates shows how much Ethernet frames can cross
the wireless link under optimal condition. However, MAC
rates are dependent on PHY Rates and can give us fair idea
of how TCP throughput varies. Formula for MAC rates:
\begin{equation}
MAC_T = PHY_T * Cycle_T * Efficiency_T 
\end{equation}
Where:
Tx MAC Rates = MAC\textsubscript{T} \\
Tx PHY Rates = PHY\textsubscript{T} \\
MAC Duty Cycle = Cycle\textsubscript{T} \\
Tx MAC Efficiency = Efficient\textsubscript{T} \\
PHY\textsubscript{R} $*$ Cycle\textsubscript{R} $*$ Efficiency\textsubscript{R} = MAC\textsubscript{R} \\
Rx MAC Rates = MAC\textsubscript{R} \\
Rx PHY Rates = PHY\textsubscript{R} \\
Rx MAC Duty Cycle = Cycle\textsubscript{R} \\
MAC Efficiency = Efficiency \textsubscript{R} \\

A simple iperf test is performed to collect the TCP
throughput rates. Iperf provides room to define TCP
parameters. For the sake of simplicity, we will set our iperf
parameters as defined below: 
\begin{itemize}
    \item Number of TCP Connection – 50 and TCP Window size - 64K 
\end{itemize}

\begin{figure}[t]
    \centering
    \includegraphics[width=0.65\linewidth]{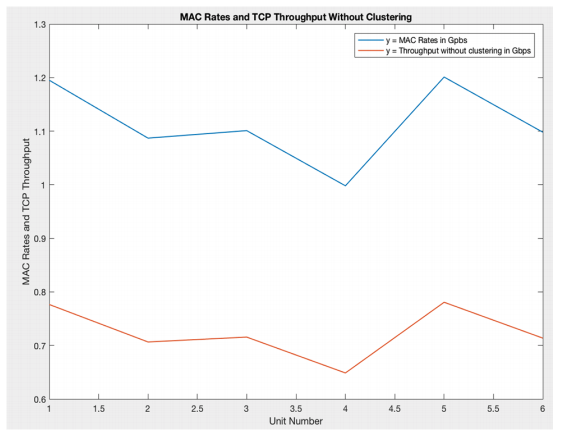}
    \captionsetup{font=footnotesize}
    \caption{MAC Rates and Throughput without
Clustering
    }
    \label{fig:MACRatesThroughputWithoutClustering}
\end{figure}

\begin{figure}[t]
    \centering
    \includegraphics[width=0.65\linewidth]{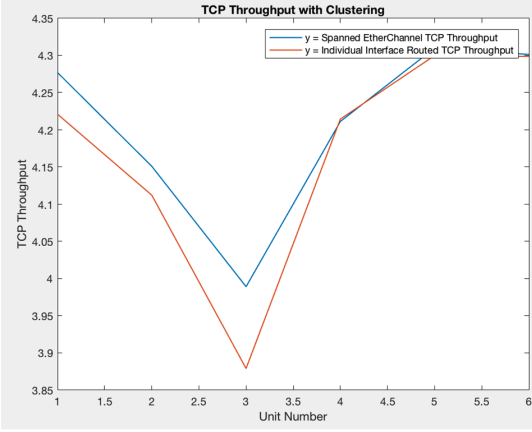}
    \captionsetup{font=footnotesize}
    \caption{Throughput With Clustering
Clustering
    }
    \label{fig:ThroughputWithClustering}
\end{figure}

\begin{figure}[t]
    \centering
    \includegraphics[width=0.65\linewidth]{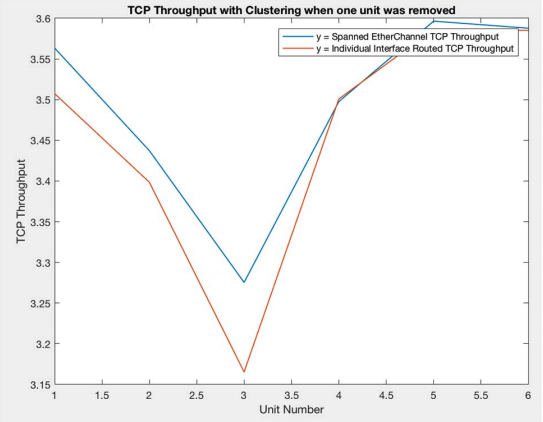}
    \captionsetup{font=footnotesize}
    \caption{Throughput with one unit removed 
    }
    \label{fig:ThroughputWithOneUnitRemoved}
\end{figure}

\begin{figure}[t]
    \centering
    \includegraphics[width=0.65\linewidth]{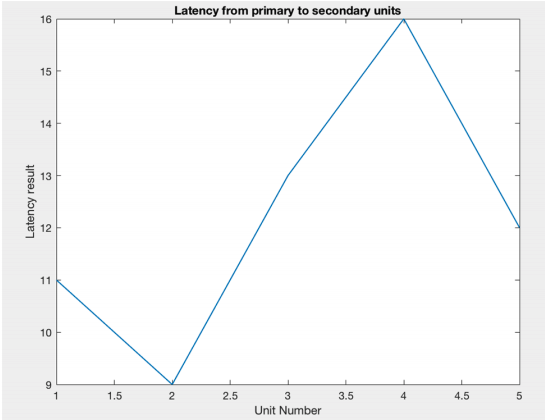}
    \captionsetup{font=footnotesize}
    \caption{ Latency from primary to secondaries.  
    }
    \label{fig:LatenciesFromPrimaryToSecondaries}
\end{figure}

\chapter{Conclusion}
With CNET APIs infrastructure, and changing duty cycle and efficiency, we could simulate the wireless backhaul links.
Different scenarios were taken into account.
Figures demonstrated that throughput increased. 
Additionally, one unit was brought down and we could see the effect on throughput.
 In figure \ref{fig:MACRatesThroughputWithoutClustering}, MAC rates and throughput are shown. This
is without clustering and units are working independently
and transmission. These PHY rates are being used in the
rest of the test cases we captured. Individual units give
throughput around 0.8 to 1.3 Mbps. In figure \ref{fig:ThroughputWithClustering}, the throughput is with clustering when all units are working
together and projecting as a single logical unit. Figure \ref{fig:ThroughputWithOneUnitRemoved}
shows throughput when a single unit is removed. Figure \ref{fig:LatenciesFromPrimaryToSecondaries} results shows that throughput got increased
and latency was under 20 ms from primary to secondaries.
When we have one unit removed/failed the throughput got
decreased considerable.
Throughput shown in the result proves that with
clustering we have more resilient, robust, and highly
available solution for providing higher throughput. Even if
we have one unit removed, throughput doesn’t decrease
significantly

\pagebreak

\nocite{*} 
\bibliographystyle{unsrt}
\bibliography{main} 
\addcontentsline{toc}{chapter}{\bibname}

\end{document}